\documentclass[floatfix, noshowpacs, preprintnumbers, twocolumn, amsmath, amssymb, aps, superscriptaddress, prl]{revtex4}

\pdfoutput=1

\usepackage{bm}
\usepackage{amsmath}
\usepackage{amssymb}
\usepackage{graphicx}
\usepackage{xcolor}
\usepackage{bbold}
\usepackage{url}
\usepackage[caption=false]{subfig}

\begin{document}

\title{Efficient experimental validation of photonic boson sampling against the uniform distribution}

\author{Nicol\`o Spagnolo}
\affiliation{Dipartimento di Fisica, Sapienza Universit\`{a} di Roma,
Piazzale Aldo Moro 5, I-00185 Roma, Italy}

\author{Chiara Vitelli}
\affiliation{Dipartimento di Fisica, Sapienza Universit\`{a} di Roma,
Piazzale Aldo Moro 5, I-00185 Roma, Italy}
\affiliation{Center of Life NanoScience @ La Sapienza, Istituto
Italiano di Tecnologia, Viale Regina Elena, 255, I-00185 Roma, Italy}

\author{Marco Bentivegna}
\affiliation{Dipartimento di Fisica, Sapienza Universit\`{a} di Roma,
Piazzale Aldo Moro 5, I-00185 Roma, Italy}

\author{Daniel J. Brod}
\affiliation{Instituto de F\'isica, Universidade Federal Fluminense, Av. Gal. Milton Tavares de Souza s/n, Niter\'oi, RJ, 24210-340, Brazil}

\author{Andrea Crespi}
\affiliation{Istituto di Fotonica e Nanotecnologie, Consiglio
Nazionale delle Ricerche (IFN-CNR), Piazza Leonardo da Vinci, 32,
I-20133 Milano, Italy}
\affiliation{Dipartimento di Fisica, Politecnico di Milano, Piazza
Leonardo da Vinci, 32, I-20133 Milano, Italy}

\author{Fulvio Flamini}
\affiliation{Dipartimento di Fisica, Sapienza Universit\`{a} di Roma,
Piazzale Aldo Moro 5, I-00185 Roma, Italy}

\author{Sandro Giacomini}
\affiliation{Dipartimento di Fisica, Sapienza Universit\`{a} di Roma,
Piazzale Aldo Moro 5, I-00185 Roma, Italy}

\author{Giorgio Milani}
\affiliation{Dipartimento di Fisica, Sapienza Universit\`{a} di Roma,
Piazzale Aldo Moro 5, I-00185 Roma, Italy}

\author{Roberta Ramponi}
\affiliation{Istituto di Fotonica e Nanotecnologie, Consiglio
Nazionale delle Ricerche (IFN-CNR), Piazza Leonardo da Vinci, 32,
I-20133 Milano, Italy}
\affiliation{Dipartimento di Fisica, Politecnico di Milano, Piazza
Leonardo da Vinci, 32, I-20133 Milano, Italy}

\author{Paolo Mataloni}
\affiliation{Dipartimento di Fisica, Sapienza Universit\`{a} di Roma,
Piazzale Aldo Moro 5, I-00185 Roma, Italy}
\affiliation{Istituto Nazionale di Ottica (INO-CNR), Largo E. Fermi 6, I-50125 Firenze, Italy}

\author{Roberto Osellame}
\email{roberto.osellame@polimi.it}
\affiliation{Istituto di Fotonica e Nanotecnologie, Consiglio
Nazionale delle Ricerche (IFN-CNR), Piazza Leonardo da Vinci, 32,
I-20133 Milano, Italy}
\affiliation{Dipartimento di Fisica, Politecnico di Milano, Piazza
Leonardo da Vinci, 32, I-20133 Milano, Italy}

\author{Ernesto F. Galv\~{a}o}
\email{ernesto@if.uff.br}
\affiliation{Instituto de F\'isica, Universidade Federal Fluminense, Av. Gal. Milton Tavares de Souza s/n, Niter\'oi, RJ, 24210-340, Brazil}

\author{Fabio Sciarrino}
\email{fabio.sciarrino@uniroma1.it}
\affiliation{Dipartimento di Fisica, Sapienza Universit\`{a} di Roma,
Piazzale Aldo Moro 5, I-00185 Roma, Italy}
\affiliation{Istituto Nazionale di Ottica (INO-CNR), Largo E. Fermi 6, I-50125 Firenze, Italy}

\maketitle

\textbf{A boson sampling device is a specialised quantum computer that solves a problem which is strongly believed to be computationally hard for classical computers \cite{Aaronson10}. Recently a number of small-scale implementations have been reported \cite{Crespi2012,Till2012,Broome2013,Spring2013}, all based on multi-photon interference in multimode interferometers. In the hard-to-simulate regime, even validating the device's functioning may pose a problem \cite{Barz2013}. In a recent paper, Gogolin \textit{et al.} \cite{Gogolin2013} showed that so-called symmetric algorithms would be unable to distinguish the experimental distribution from the trivial, uniform distribution. Here we report new boson sampling experiments on larger photonic chips, and analyse the data using a scalable statistical test recently proposed by Aaronson and Arkhipov \cite{Aaronson13}. We show the test successfully validates small experimental data samples against the hypothesis that they are uniformly distributed. We also show how to discriminate data arising from either indistinguishable or distinguishable photons. Our results pave the way towards larger boson sampling experiments whose functioning, despite being non-trivial to simulate, can be certified against alternative hypotheses.} 

Large-scale quantum computers hold the promise of efficiently solving problems which are believed to be intractable for classical computers, such as integer factoring \cite{Shor97}. We are, however, far from being able to experimentally demonstrate a large-scale, universal quantum computer \cite{Ladd2010}. This has motivated the recent study of different classes of restricted quantum computers \cite{Barreiro2011,Islam2013}, which may provide a more feasible way of experimentally establishing what has been called the quantum computational supremacy \cite{Preskill12} over classical computers.

\begin{figure*}[ht!]
\includegraphics[width=0.84\textwidth]{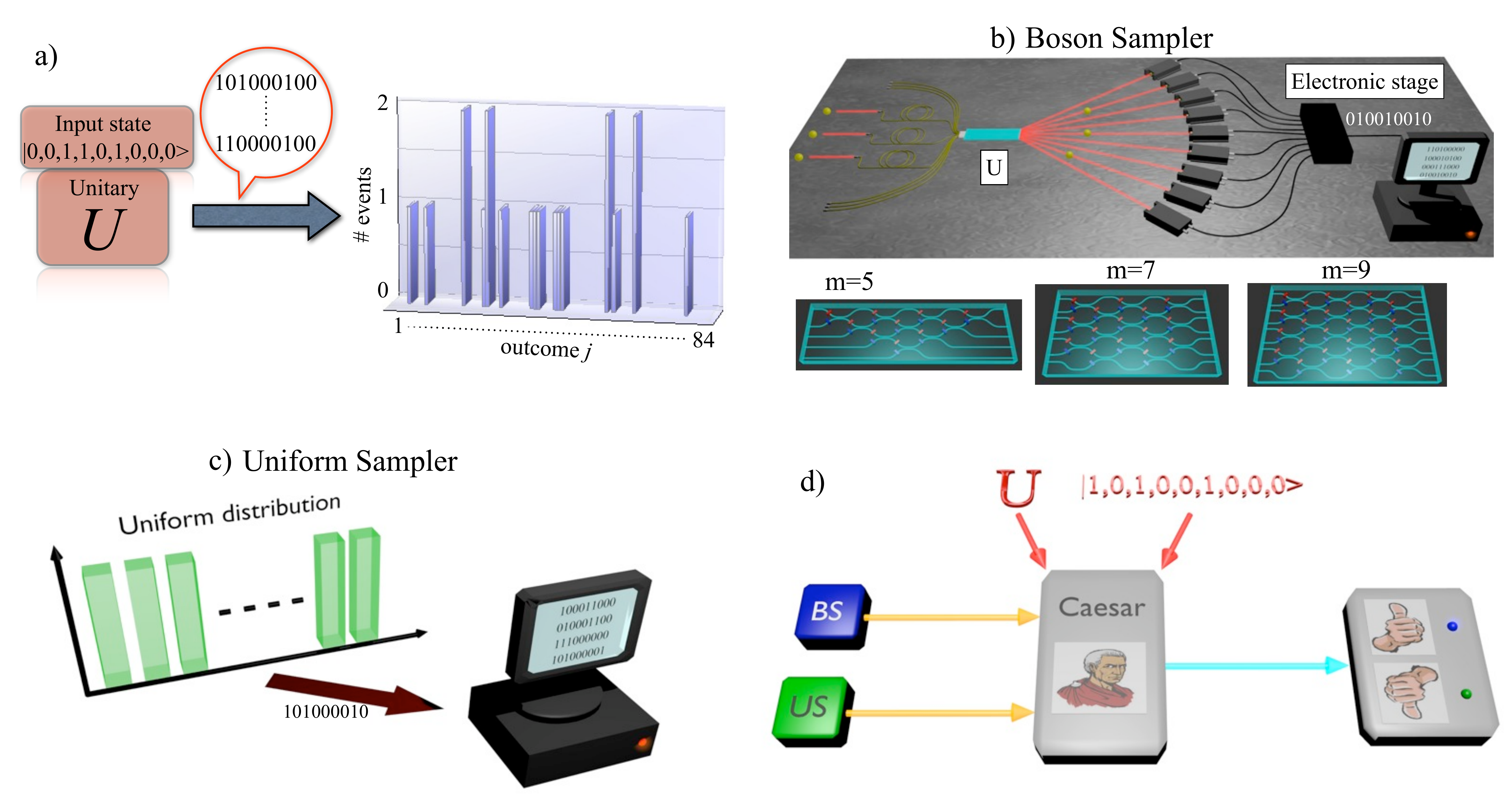}
\caption{\textbf{Boson sampling and its certification.} a) The Boson Sampling problem consists in sampling from the output distribution of $n$ bosons evolving according to a linear transformation $U$. The $m\times m$ unitary matrix $U$ together with the input state are known quantities in the problem. b) Photonic implementation of Boson Sampling: $n$ indistinguishable photons interfere in a random, linear $m$-mode interferometer, with photo-detection at the output modes. Let us call Boson Sampler an agent that provides events generated by a Boson Sampling experiment; our implementation is shown schematically here.  c) The Uniform Sampler is an agent that generates events classically according to a uniform distribution over outputs. d) We are interested in how well a third agent, the certifier Caesar, can use information about $U$ to distinguish the output data sets generated by Boson Sampler from those generated by Uniform Sampler.}
\label{fig:task}
\end{figure*}

One example of these restricted quantum computers are multi-mode interferometers designed to solve the Boson Sampling problem \cite{Aaronson10}, recently demonstrated in small-scale photonic experiments \cite{Crespi2012,Till2012,Broome2013,Spring2013}. The Boson Sampling problem involves simulating the following quantum experiment (see Fig. \ref{fig:task} a,b): input $n$ bosons in different modes of an $m$-mode linear interferometer ($m>n$) and measure the distribution of bosons at the interferometer's output modes. If performed with indistinguishable bosons, this experiment results in an output distribution which is hard to sample, even approximately, on classical computers \cite{Aaronson10} (under very mild computational complexity assumptions). The input for the classical simulation consists in the $m \times m$ unitary matrix $U$ describing the interferometer and the list of $n$ input modes used. It is desirable to choose $U$ randomly, both to avoid regularities that could simplify the classical simulation, and because the main hardness-of-simulation proof of \cite{Aaronson10} holds only for uniformly sampled unitaries. These recent theoretical and experimental results motivated further investigations on error tolerances \cite{Rohde12,Leverrier2013}, as well as additional analyses of optical implementations \cite{Motes2013,Rohde2013}. Very recently, there appeared a proposal to implement Boson Sampling computers using trapped ions \cite{Shen2013}.

It has been suggested recently, however, that due to their very complexity, large boson sampling experiments could not possibly be validated, i.e. their proper functioning could not be ascertained. Gogolin \textit{et al.} showed in \cite{Gogolin2013} that so-called symmetric algorithms fail to distinguish the distribution of experimental data from the trivial, uniform distribution. Intuitively, it seems hard to use an experimental data set of polynomial size (in $n$) to distinguish two distributions over a sample space which is exponentially large. This criticism put in question the notion that larger boson sampling experiments could be shown to decisively outperform classical computers. 
 
This criticism of Boson Sampling has been recently refuted by Aaronson and Arkhipov \cite{Aaronson13}, who argued that it was unreasonable to restrict the statistical analysis to symmetric algorithms only. Moreover, Aaronson and Arkhipov showed that it is indeed possible to discriminate experimental data against the uniform distribution by taking advantage of the input data of the boson sampling problem (the unitary $U$ and the input state). They proposed a scalable validation test which, for large enough $n$ and uniformly-drawn unitaries, succeeds with high probability using only a constant number of samples (see the Methods section for a description of the test).

Here we report photonic boson sampling experiments performed with $3$ photons in randomly-designed integrated chips with $5,7$ and $9$ modes, corresponding to $10,35$ and $84$ different no-collision outputs, i.e. outputs with at most one photon per mode. We analyse the experimental data using the Aaronson-Arkhipov validation test \cite{Aaronson13}, showing that the test works in practice, even in the presence of experimental imperfections. We also show how we can successfully tell apart data corresponding to distinguishable and indistinguishable photons in these experiments.

\begin{figure*}[ht!]
\includegraphics[width=0.85\textwidth]{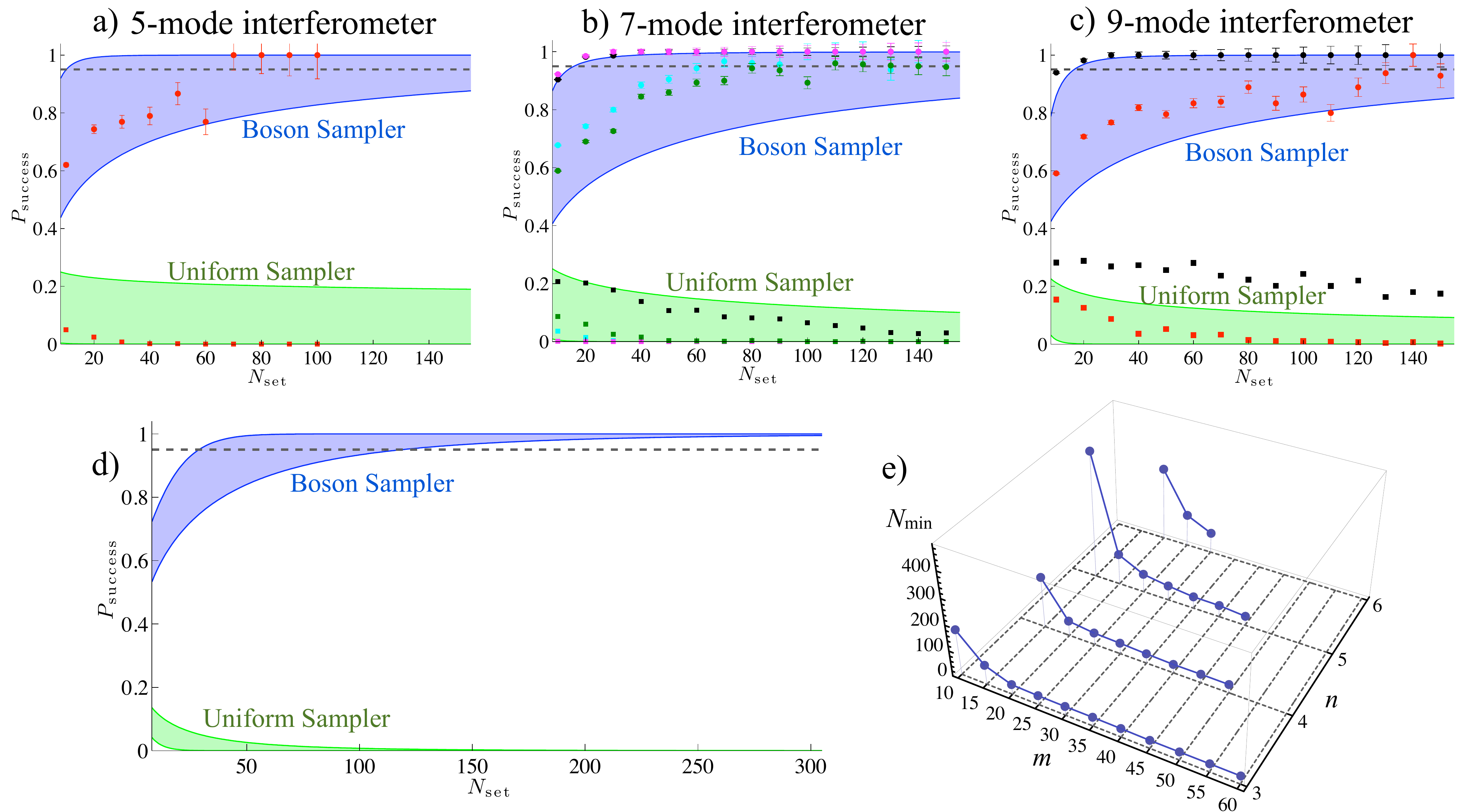}
\caption{\textbf{Experimental validation of boson sampling.} Performance of the validation test of reference \cite{Aaronson13} using experimental data sets of varying sizes. Here we show Caesar's success rate $P_{\mathrm{success}}$ in distinguishing the sets, as a function of set size $N_{\mathrm{set}}$, in experiments using a) one Haar-random 5-mode interferometer (red circle points: input state $\vert 0,1,1,1,0\rangle$); b) four different three-photon input combinations in a random 7-mode interferometer (green circle points: input state $\vert 0,0,1,1,1,0,0 \rangle$; cyan circle points: input state $\vert 0,0,0,1,1,1,0 \rangle$; black circle points: input state $\vert 0,1,0,1,0,1,0 \rangle$; magenta circle points: input state $\vert 1,1,1,0,0,0,0 \rangle$); c) two different three-photon inputs in a random 9-mode interferometer (red circle points: input state $\vert 0,0,0,1,1,1,0,0,0 \rangle$; black circle points: input state $\vert 0,0,1,1,0,0,0,1,0 \rangle$). Grey dashed line: level for $95\%$ success probability. Square points: numerical simulations, averaged over 1000 data sets of size $N_{\mathrm{set}}$, of the validation test for data generated by Uniform Sampler. In all plots, blue shaded region corresponds to the theoretical prediction of the Boson sampling validation, reported as $1.5$-standard deviation and obtained by averaging over a numerical simulation with 1000 Haar-uniform unitaries. The average is performed by excluding the unitaries where the success rate does not reach $95\%$ even with $N_{\mathrm{set}}=5000$. The number of unitaries with the (asymptotically proven) correct behaviour was respectively $434$ ($m=5$), 573 ($m=7$), 822 ($m=9$) and, for fixed $n$, increases with $m$. Green shaded region corresponds to the theoretical prediction for the validation of a uniform sampler, reported as $1.5$-standard deviation over a numerical simulation with 1000 Haar-uniform unitaries. d) Simulated performance of the validation test for Boson Sampling experiments with $n=5$ photons in 100 Haar-uniform 25-mode interferometers, plotting the $1.5$-standard deviation average success rate obtained. e) Minimum data set size $N_{\mathrm{min}}$ to obtain $>95\%$ success probability for Boson Sampling experiments and to obtain $<5\%$ success probability for Uniformly-sampled experiments, as a function of the number of photons $n$ and of the number of modes $m$ obtained through a numerical simulation. For each point, the simulation is averaged over 50 or 100 Haar-uniform unitaries.}
\label{fig:analysis}
\end{figure*}

To perform boson sampling experiments we used three ingredients: a three-photon source, randomly designed interferometers and a detection apparatus able to record all the three-photon coincidence events at the output of the interferometer (see Fig. \ref{fig:task} b). The three-photon input state is produced by exploiting the second order parametric down conversion process, with three photons sent into the interferometers and a fourth one used as a trigger. We fabricated stable, integrated \cite{Poli2008} interferometers with $5,7$ and $9$ modes in a glass chip by femtosecond laser waveguide writing \cite{Davis96,Osellame2003,gattass2008flm}. This technique consists in a direct inscription of waveguides in the glass volume, exploiting the nonlinear absorption of focused femtosecond laser pulses to induce a permanent and localized increase in the refractive index. Single photons may jump between waveguides by evanescent coupling in regions where waveguides are brought close together, thus realising the beam splitter transformation. Precise control of the coupling between the waveguides and of the photon path lengths, enabled by a 3D waveguide design \cite{Crespi2012}, allowed us to engineer arbitrary interferometers by cascading directional couplers and phase shifters with different layouts (see Fig. \ref{fig:task} b and Refs. \cite{Sans2012,Crespi2013,Spag2012,Crespi2012,Spagnolo2013}). Finally, single photon counting detectors and an electronic acquisition apparatus are used to reconstruct the probabilities associated with all three-photon coincidence events at the chip's output. Further details on the integrated circuits and on the experimental set-up are given in the Methods section.

Let us now discuss how a certifier (Caesar) can validate small sets of boson sampling data generated by an agent we call the Boson Sampler (BS), against the hypothesis that they might have been generated by Uniform Sampler (US), an agent that samples from the uniform probability distribution (see Fig. \ref{fig:task} b-d). Caesar succeeds by using the Aaronson-Arkhipov test \cite{Aaronson13} (described in the Methods section) on small experimental data sets. We have applied this test to multiple, random experimental data sets of varying sizes. This enabled us to gauge the trade-off between set size and success rate, which has been theoretically proven to be high only for large enough $n$ \cite{Aaronson13}. The results are shown in Fig. \ref{fig:analysis}. For the experiments with the $5$-, $7$-, and $9$-mode chips, Caesar reaches a $95\%$ average success rate with very modest set sizes of just $\sim 100$ events. This establishes experimentally the usefulness of the Aaronson-Arkhipov test \cite{Aaronson13} for the analysis of small-scale experiments.

\begin{figure*}[ht!]
\includegraphics[width=0.85\textwidth]{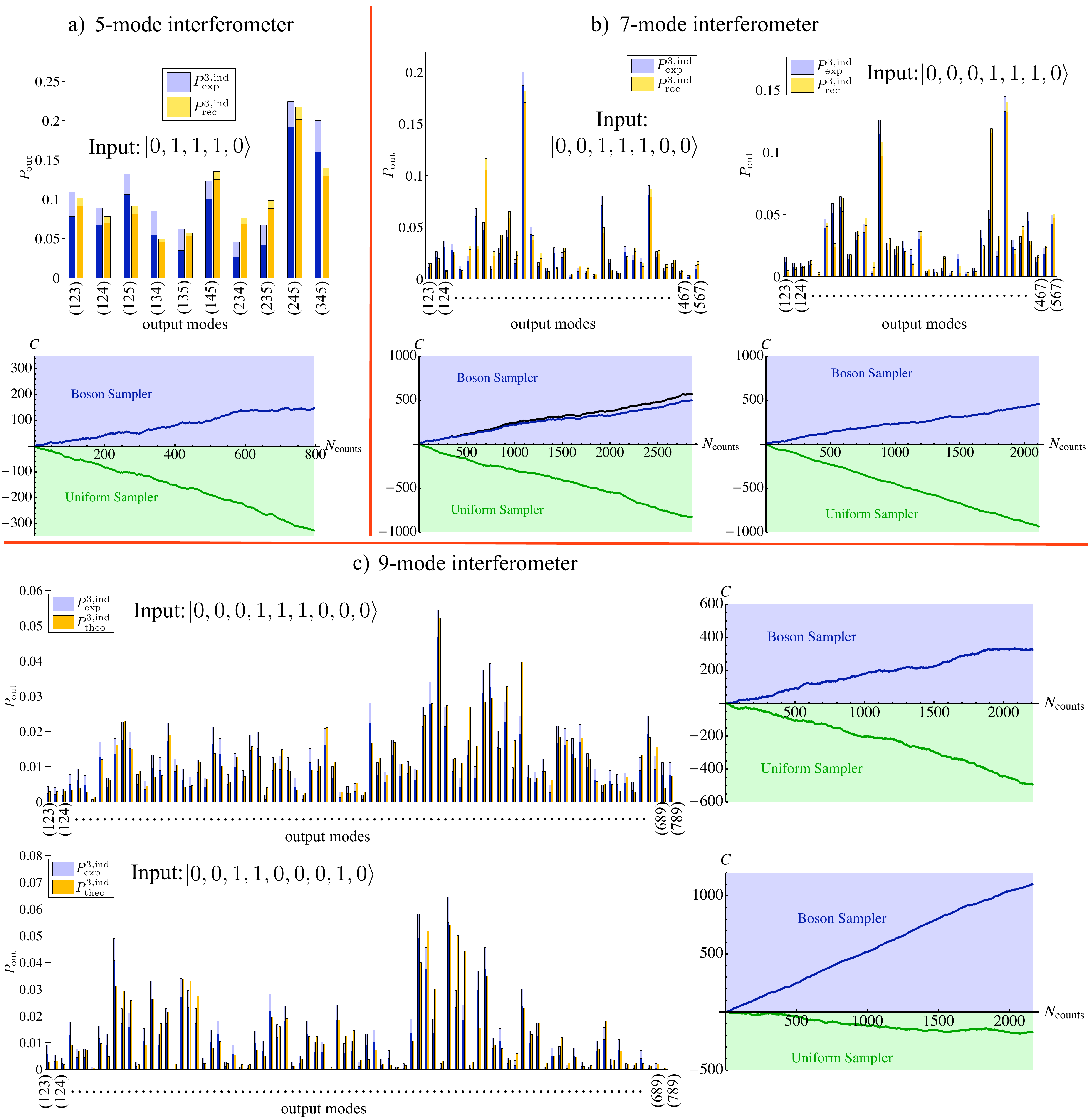}
\caption{\textbf{Full validation of the Boson Sampling experiments.} Here we compare the experimentally measured probabilities $P_{\mathrm{out}}$ of all no-collision outputs of our Boson Sampling experiments with the expected probabilities (yellow bars) for a) Haar-random 5-mode chip with input state $\vert 0,1,1,1,0 \rangle$; b) random 7-mode chip with input states $\vert 0,0,1,1,1,0,0 \rangle$ and $\vert 0,0,0,1,1,1,0 \rangle$; c) random 9-mode chip with input states $\vert 0,0,0,1,1,1,0,0,0 \rangle$ and $\vert 0,0,1,1,0,0,0,1,0 \rangle$. The expected probabilities take into account the partial photon distinguishability of the source and multiphoton events (see Methods and Ref. \cite{Crespi2012,Spagnolo2013}). Lighter regions of the blue bars correspond to the experimental error, which is due to the poissonian statistics of the events. Lighter regions of the yellow bars correspond to the error in the reconstruction process, retrieved by a MonteCarlo simulation. Bottom figures of panels a) and b), right figures of panel c): application of the Aaronson-Arkhipov test to the full set of experimental data. $C$ is a counting variable that is increased by 1 for each event assigned to the Boson Sampler, and decreased by 1 for each event assigned to the Uniform Sampler. When $C>0$, the complete data set is assigned to the Boson Sampler. Blue points: test applied on the experimental data by exploiting the ideal unitary $U_t$. Black points: test applied on the experimental data by exploiting the reconstructed unitary $U_r$ (for the $m=5,7$ chips only). Green points: test applied on simulated data generated by the Uniform Sampler. For states $\vert 0,1,1,1,0 \rangle$ ($m=5$) and $\vert 0,0,0,1,1,1,0 \rangle$ ($m=7$) blue and black points present a large overlap and superimpose in the figures.}
\label{fig:histogram}
\end{figure*}

To show the test will also work in as-yet unperformed, larger-scale experiments, in Fig. \ref{fig:analysis}d we find the test's success rate for simulated boson sampling experiments with $n=5$ and $m=20$. Additionally, in Fig. \ref{fig:analysis}e we numerically determine the minimum data set size $N_{\mathrm{min}}$ for which the Aaronson-Arkhipov test discriminated boson sampling data from the uniform distribution (and viceversa) with a success rate $>95\%$. Not only is $N_{\mathrm{min}}$ small for all experiments we simulated, it actually decreases as we increase $m$. Despite working for all interferometers we implemented experimentally, our numerical simulations revealed that the test fails for some interferometers if the ratio $m/n$ is too low.

\begin{figure*}[ht!]
\centering
\includegraphics[width=0.7\textwidth]{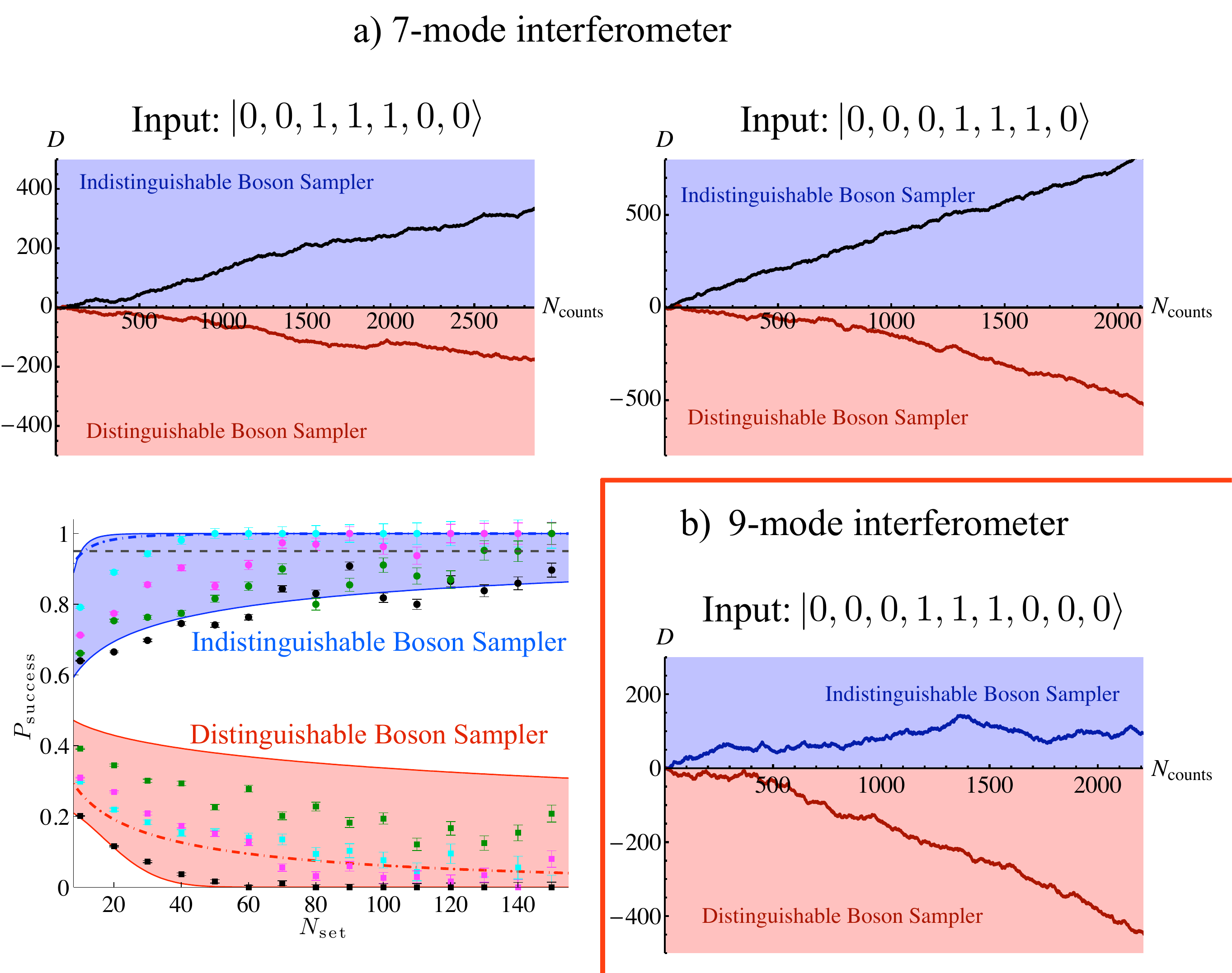}
\caption{\textbf{Discrimination between alternative distributions}. a) Experimental results of the discrimination between boson sampling distributions with distinguishable or indistinguishable photons for the 7-mode chip. The protocol is applied by using the probability distributions obtained from the reconstructed unitary $U_{r}$. Top figures: evaluation of the $D$ parameter for two different input states. Black data: indistinguishable photons. Red data: distinguishable photons. Bottom figure: success probability $P_{\mathrm{success}}$ of the discrimination protocol as a function of the data set size $N_{\mathrm{set}}$. Green circle points: input state $\vert 0,0,1,1,1,0,0 \rangle$. Blue circle points: input state $\vert 0,1,0,1,0,1,0 \rangle$. Cyan circle points: input state $\vert 0,0,0,1,1,1,0 \rangle$. Magenta circle points: input state $\vert 1,1,1,0,0,0,0 \rangle$. Square points: corresponding success probability of the protocol for the ``false'' positive events with distinguishable photons. Blue shaded region: numerical simulation of the success probability of discrimination test for indistinguishable photons, taking into account the partial photon distinguishability of the adopted source (see Methods). The results are averaged over 1000 Haar-uniform unitaries, and are reported as $1.5$-standard deviation. Blue dash-dotted line: average behaviour for perfectly indistinguishable photons. Red shaded region: numerical simulation for distinguishable photons, reported as $1.5$-standard deviation, where the partial photon distinguishability of the adopted source has been taken into account in the indistinguishable photon distribution. Red dash-dotted line: average behaviour without taking into account partial photon-distinguishability. b) Experimental results of the discrimination between boson sampling distributions with distinguishable or indistinguishable or photons for the 9-mode chip with input state $\vert 0,0,0,1,1,1,0,0,0 \rangle$. The protocol is applied by using the probability distributions obtained from the ideal unitary $U_{t}$. Blue data: indistinguishable photons. Red data: distinguishable photons.}
\label{fig:indvsdis}
\end{figure*}

In the probed regime with $n=3$ photons and interferometers with up to $m=9$ modes it is possible to perform a full validation of the boson sampling experiments by reconstructing all probabilities associated with no-collision events. This requires recording experimental data sets of a larger size; for the $m=7$ chip, for example, we recorded $\sim 2100$ events. The experimentally reconstructed probabilities are then compared with the theoretical prediction, obtained by applying the permanent formula \cite{Aaronson10}. For the chips with $m=5,7$, we fully reconstructed the interferometer unitaries using single-photon and two-photon experiments \cite{Crespi2012,Obrien12}, while for the $m=9$ we compared the three-photon data with what is expected from the ideal, theoretical unitary $U_t$. The results are shown in Fig. \ref{fig:histogram}, and the good agreement between the experiments and the predictions is quantified by the variation distance $d=1/2 \sum_{k} \vert p_k - q_k \vert$, which reaches values $d_{\mathrm{exp,r}}^{(2,3,4)}= 0.104 \pm 0.022$ ($m=5$), $d_{\mathrm{exp,r}}^{(3,4,5)}=0.168 \pm 0.016$ and $d_{\mathrm{exp,r}}^{(4,5,6)}=0.133 \pm 0.017$ ($m=7$), $d_{\mathrm{exp,t}}^{(4,5,6)}=0.113 \pm 0.017$  and $d_{\mathrm{exp,t}}^{(3,4,8)}=0.167 \pm 0.020$ ($m=9$). Furthermore, we have applied the Aaronson-Arkhipov test \cite{Aaronson13} to the full data set (Fig. \ref{fig:histogram}). We observe that this test can be applied successfully by using either the reconstructed unitary $U_r$ or the ideal unitary $U_t$.

In addition, simple tests can be used to validate boson sampling data against probability distributions which are more similar to the experimental data than the uniform distribution is. In the Methods section we describe such a test, which is a standard likelihood ratio test \cite{CoverThomas06} with added thresholds to better take into consideration experimental imperfections. We have applied the test to discriminate boson sampling experiments performed with either indistinguishable or distinguishable photons in a 7-mode and a 9-mode interferometer (see Fig. \ref{fig:indvsdis}). The regime of distinguishable photons has been obtained experimentally by introducing a relative temporal delay between the three photons larger than their coherence time. Note that the test requires calculating the probabilities associated with each observed outcome, and hence it is not computationally efficient. A laptop can, however, test a small sample of boson sampling data with up to about $n=25$ photons in a day or so. No boson sampling simulation algorithm has yet been developed that is significantly faster than computing the full output probability distribution (which involves calculating an exponential number of permanents) \cite{Aaronson13b}. While establishing better bounds on the complexity of simulation is an important open theoretical problem, in the absence of further progress there is a \textit{de facto} exponential gap between the (known) complexities of simulating boson sampling and validating experimental data against various alternative probability distributions.  This simple observation could, in the near future, guide the design of boson sampling experiments that are computationally non-trivial to simulate on a classical computer, but still feasible to validate.

Our experiments have shown how to leverage available information about the boson sampling experiment to distinguish experimental data from the uniform distribution, using the scalable test proposed by Aaronson and Arkhipov \cite{Aaronson13}. Our analysis shows that this test works even for small instances of boson sampling experiments, and provides experimental support for the recent theoretical refutation \cite{Aaronson13} of a recent criticism of boson sampling experiments \cite{Gogolin2013}. We have also certified the experimental data using a test that distinguishes them from a similar experiment done with distinguishable photons. Our results show the feasibility of certifying boson sampling experiments against various alternative hypotheses. This will be decisive in future experiments that use boson sampling devices to establish the quantum information processing supremacy over classical computers.

\textbf{Acknowledgements.} We acknowledge very useful feedback from S. Aaronson, L. Aolita and A. Arkhipov. This work was supported by the ERC-Starting Grant 3D-QUEST (3D-Quantum Integrated Optical Simulation; grant agreement no. 307783): http://www.3dquest.eu, and by the Brazilian National Institute for Science and Technology of Quantum Information (INCT-IQ/CNPq).

\section*{Methods}

\textbf{Fabrication.} Multimode integrated interferometers were fabricated in glass chips by femtosecond laser writing \cite{gattass2008flm,Osellame2003} with layout as in Ref. \cite{Crespi2012,Spagnolo2013}. To inscribe the waveguides, laser pulses with 220 nJ energy and 1 MHz repetition rate from an Yb:KYW cavity dumped oscillator were focused through a 0.6 NA microscope objective 170 $\mu$m under the sample surface. The laser pulses are about 300 fs long and have 1030 nm wavelength. The sample was translated at constant speed, drawing in the three dimensions the desired waveguide paths into the boro-aluminosilicate glass (EAGLE2000, Corning Inc.). The fabricated waveguides yield single mode behaviour at 800 nm wavelength, with about 0.5 dB/cm propagation losses.

The architecture of the interferometers are shown in Fig. \ref{fig:task}b. The $m=5$ device corresponds to a Haar-random unitary, and has been implemented by decomposing the unitary in beam-splitter operations and phase-shifters according to the procedure shown in Ref. \cite{Reck94}. The $m=7,9$ interferometers are obtained by drawing a set of random phases, and by implementing the corresponding network with balanced 50/50 directional couplers.

\textbf{Experimental setup.}
Four photons were produced in the pulsed regime at $785$ nm exploiting the second order  parametric down conversion process, by pumping a 2 mm long BBO crystal by the $392.5$ nm wavelength pump field. Typical count rates of the source were around $250000$ Hz for the four signals, $40000$ Hz for the two-fold coincidences and $20$ Hz for the four-fold coincidences. One of the photons, adopted as a trigger, was filtered by $3$ nm interferential filters, coupled into a single-mode fiber and detected by a single-photon counting detector. For the other three photons, spectral filtering by $3$ nm interferential filters, coupling into single-mode fibers, polarization compensation, and propagation through different delay lines were performed before coupling into the chips. At the output of the chip multimode fibers were connected to single photon counting detectors. The four-fold coincidences between the three-photon state and the trigger signal were acquired by an electronic system: a timing circuit driven by the trigger signal controlled both the conditioning input circuit and the four-fold coincidences discrimination and detection circuit. The circuit generates the handshake signal sent to the National PCI-6503 board used for recognising the pattern.

Due to the spectral correlations of the generated photon pairs, the emitted photons present a degree of partial distinguishability (see Refs. \cite{Crespi2012,Spagnolo2013}). Furthermore, down-conversion sources present a non-zero probability to generate six-photon events. These effects have been taken into account in the expected probability distributions of Fig. \ref{fig:histogram}.

\textbf{The validation test of Aaronson and Arkhipov \cite{Aaronson13}.} The validation test of Aaronson and Arkhipov \cite{Aaronson13} works as follows. Let us assume that the $n$ input photons occupy the set of modes $S=\{s_1, s_2, ..., s_n \}$ of the $m$-mode interferometer, which is described by a $m \times m$ unitary matrix $U$. Each single experimental outcome consists of photons leaving the interferometer in a set of modes $T=\{t_1, t_2, ..., t_n \}$, where we have assumed an experiment which only detects no-collision events. Define a $n\times n$ submatrix $A$ of $U$ with elements $A_{i,j}=U_{s_i, t_j}$. Now calculate $P=\prod_{i=1}^n \sum_{j=1}^n |A_{i,j}|^2$. If $P>\left(\frac{n}{m}\right)^n$, guess the outcome is from the boson sampling distribution, otherwise guess the outcome arose from a uniform distribution over the possible outputs. Notably, the test is computationally efficient, as it does not involve the calculation of any permanents. Assuming the interferometer's unitary is picked from the uniform, Haar distribution, this test has been proven to succeed with probability $1-O(\delta)$ and for sufficiently large $n$, provided  $m>n^{5.1}/\delta$ \cite{Aaronson13}. To increase the success rate, repeat the process with a sample of $N$ experimental outcomes and use majority voting to decide which case is more likely to hold. 

\textbf{Validating boson sampling data against distinguishable photon distribution.} Here we describe a test that can validate boson sampling data against the hypothesis that the photons are distinguishable. The test is an adapted version of the likelihood ratio test \cite{CoverThomas06} which incorporates a discrimination threshold to compensate for experimental noise. Let $p^{\mathrm{ind}}_{i}$ and $q^{\mathrm{dis}}_{i}$ be the probabilities associated with indistinguishable and distinguishable photons for the measured outcome, and let $D$ be the discrimination parameter, initialized to the value $D=0$. For each experimental outcome, we calculate the ratio of the expected probabilities for indistinguishable and distinguishable photons. If the ratio is close to one, up to to a threshold  $k_{1} < p^{\mathrm{ind}}_{i}/q^{\mathrm{dis}}_{i} < 1/k_{1}$, the event is considered to be inconclusive and $D$ is left unchanged. These inconclusive events, however, are still counted as a resource and do contribute to the effective number of events required to discriminate the two distributions. If $1/k_{1} \leq p^{\mathrm{ind}}_{i}/q^{\mathrm{dis}}_{i} < k_{2}$, the event is assigned to the Indistinguishable Boson Sampler by adding $+1$ to $D$. If the ratio between the two probabilities is high, $p^{\mathrm{ind}}_{i}/q^{\mathrm{dis}}_{i} \geq k_{2}$, the event is assigned to the Indistinguishable Boson Sampler by adding $+2$ to $D$, thus reflecting the higher level of confidence in this case. Conversely, if $1/k_{2} < p^{\mathrm{ind}}_{i}/q^{\mathrm{dis}}_{i} \leq k_{1}$ and $p^{\mathrm{ind}}_{i}/q^{\mathrm{dis}}_{i} \leq 1/k_{2}$ the event is assigned to the Distinguishable Boson Sampler by adding $-1$ and $-2$ to $D$ respectively. Finally, after $N$ experimental outcomes, if $D>0$ the whole data set is assigned to the Indistinguishable Boson Sampler, and conversely if $D<0$. In our analysis we set $k_{1} = 0.9$ and $k_{2} = 1.5$.

\end{document}